\documentclass[twocolumn,showpacs,prl]{revtex4-1}
\usepackage{amsmath}
\usepackage{graphicx}

\usepackage{color}
\usepackage{algorithm2e}

\begin{document}
\vspace*{0.35in}

\title{Emergence of self-reinforcing information bottlenecks in multilevel selection}

\author{Cameron Smith}
\affiliation{Department of Systems and Computational Biology, Albert Einstein College of Medicine, 1301 Morris Park Ave, Bronx, NY 10461, USA}
\author{Matthieu Laneuville}
\author{Nicholas Guttenberg}
\affiliation{Earth-Life Science Institute, Tokyo Institute of Technology, Meguro, Tokyo 152-8551, Japan}

\begin{abstract}

We explain how hierarchical organization of biological systems emerges naturally during evolution, through a transition in the units of individuality.  We will  show how these transitions are the result of competing selective forces operating at different levels of organization, each level having different units of individuality.  Such a transition represents a singular point in the evolutionary process, which we will show corresponds to a phase transition in the way information is encoded, with the formation of self-reinforcing information bottlenecks.  We present an abstract model for characterizing these transitions that is quite general, applicable to many different versions of such transitions.  As a concrete example, we consider the transition to multicellularity.  Specifically, we study a stochastic model where isolated communities of interacting individuals (e.g. cells) undergo a transition to higher-order individuality (e.g. multicellularity). This transition is indicated by the marked decrease in the number of cells utilized to generate new communities from pre-existing ones. In this sense, the community begins to reproduce as a whole via a decreasing number of cells. We show that the fitness barrier to this transition is strongly reduced by horizontal gene transfer.  These features capture two of the most prominent aspects of the transition to multicellularity: the evolution of a developmental process and reproduction through a unicellular bottleneck.

\end{abstract}

\maketitle

\section*{Introduction}

Biological organisms have structure at many different levels of organization: from the scale of small molecules to genes, to chromosomes, subcellular organelles in eukaryotes, the arrangement and differentiation of individual cells in multicellular organisms, and the organisms themselves in the context of ecology \cite{Bonner2001,Grosberg2007,Rokas2008}. When constructing a theory for phenomena in such systems, there is a natural set of objects to use that separates out the parts that vary from the parts that are quite regular. In essence, these objects are the individuals of that theory and it is important to understand what drives the formation of new levels of individuality \cite{Buss1987,Michod1997,Michod1996,Michod2003,Michod2003a,Michod2006d,Rainey2010a,Ispolatov2011,DeMonte2014}.

Especially in biological systems, that natural set of objects can vary from case to case --- `species' has a much clearer definition for multicellular organisms than for bacteria. These objects are not inherent aspects of the matter that biological systems are made of, or directly arising from the underlying physical laws, but instead are a consequence of biological processes and the information stored within the system that encodes them \cite{Smith1999,Bergstrom2009,Adami2012}. This information is discovered and sustained by self-organization of the biological system, via the processes of evolutionary dynamics. This suggests that by understanding how biological systems create higher levels of organization --- a new set of individuals --- we can more generally understand the mechanisms behind the emergence of higher-level structures.

This need to create robust higher-level organization appears in experimental attempts to reproduce facets of the origins of life, where a generalized understanding would help guide intuition about what could be used to enable the systems to sustain their own organization \cite{Braakman2012a}. In pursuit of artificial cells, researchers must deal with an analogue of this problem in the form of controlling the distribution of genetic material between the artificial cells. Some cells may end up with multiple copies of the genetic material, or with none \cite{uno2014evolutionary}. In chemical origins of life scenarios, the distribution of compounds associated with complex autocatalytic networks plays a similar role \cite{Vaidya2012}. These are both cases in which the system has not yet managed to produce a robust, larger-scale individual, but in which the signatures of that larger scale can be observed for short times --- they are transitional systems.

There are similar systems that occur much later in the evolutionary history of life, which can be used as inspiration for potential solutions to the problem of creating robust higher-level organization. For example, although bacteria are generally considered to be unicellular, many exhibit features such as biofilm, chain, or syncytium formation and the absence of individual growth that are more closely associated to multicellularity \cite{Shapiro1998,Claessen2014}. The distinction between the unicellular and multicellular states of organisms is not precise \cite{Buss1987,Ratcliff2011,Libby2014,Ratcliff2015}. The transition from the unicellular to the multicellular state is most clearly indicated by germ-soma differentiation and the genetic assimilation of a developmental program that results in the reproduction of community-level morphological features \cite{Michod2001,King2004,Michod2006a,Grosberg2007,Rokas2008}. That is to say, the population-level structure becomes more precisely specified by the processes by which it is produced, usually in part through a very narrow replicative bottleneck of a single cell, followed by expansion into the new individual.

The existence of a replicative bottleneck has consequences for the evolutionary dynamics in multiscale systems, and acts as a form of population-level repair mechanism. For example, in the case of protocells, using smaller vesicles tends to suppress the emergence of parasitic genetic material \cite{ichihashi2014positive}. Other work has investigated the consequences of replicative bottlenecks, but historically much of the focus has been on the role of the bottleneck as a constraint on the stability of early forms of life in which the necessary genes for survival would not have been located on a single unified genome. 

For example, \cite{Niesert1981,Niesert1987} considers the maximum number of independent genes that can be sustained in a two-layer bag-of-genes model. There, each generation is produced through random sampling, which is used to generate a constraint on pre-biotic vesicularized genetic systems.  Similarly, the stochastic corrector model \cite{Szathmary1987,Grey1995,Zintzaras2002} looks at sub-sampling of a population via division in two as a process which can import selection pressures from the group level in order to sustain genetic composition and to overcome the stringent constraints imposed by the error threshold associated with Eigen's paradox \cite{eigen1989molecular} in prebiotic systems. Later, \cite{Takeuchi2009,Zintzaras2010} built on the artificial cells models to include multi-level selection and show that novel evolutionary directions could emerge from it.

However, bag-of-genes models are examples of systems where the population structure necessary for survival cannot change. In these cases the reproductive bottleneck helps sustain the population against mutational load, but does not act as an organizational driving force. This is because the organization is fixed and cannot change. These systems cannot do anything about the fact that their functions' are all stored in separate places and must be collectively maintained. If, on the other hand, the population structure is allowed to vary on an additional level --- perhaps allowing for multiple functions on the same genetic molecule, but perhaps isolating them --- then we can ask questions about the transitions in which the number of levels of organization of the system changes. 

Our subsequent step is to consider a three-layer system in which there are functions (genes), cells (collections of genes), and communities of cells (ensembles of collections of genes). In this case, the reproductive bottleneck is not only preventing parasitism from emerging, but also provides a feedback between the way the lower-level population is organized and the higher-level replicative dynamics. Because the cells (which are the objects being transmitted through the bottleneck) may have a variable genetic structure, the pressure produced by the bottleneck encourages gene-level organization. At the same time, if the bottleneck radius (captured in what proceeds by the sporesize variable) is capable of changing as well, then there is co-evolution between these factors. This co-evolution between individual- and community-level structure, in certain parameter ranges, can encourage the emergence of higher-level multicellular structures and retain them against fluctuations. 

Developing such a bottleneck constrains the lower level of the system, and so during the formation we expect that there will be an evolutionary pressure resisting it. As recognized by previous works \cite{Buss1987}, this introduces a difficulty --- in evolving towards a multicellular organism, the individual cells must give up their replicative identity.  During reproduction through a genetic bottleneck, mutations accumulated by somatic cells are not passed on to the offspring. In a transitional organism though, mutations which would encourage `cheating' --- that is, having an independent replicative identity --- would be beneficial at the level of the individual cells and might thus inhibit the formation of a multicellular organism.

Here we investigate a model that explicitly represents the competing interests that drive the evolution of multicellularity. On the one hand, organisms may require a relatively diverse collection of functions in order to survive in a particular environment. Individual cells are thus considered to exist as part of independent communities (biofilms) that each require a certain collection of genes to be represented in order to be viable. Pressure to retain those functions arises from survival or death on the scale of the biofilms themselves. On the other hand, once functions are being provided at the community level, selection pressures on individual cell lines to retain those functions are reduced and it becomes very likely for non-functional parasitic cells to emerge and dominate the population via an error catastrophe.

\section*{Model}

\begin{figure*}
\includegraphics[width=\textwidth]{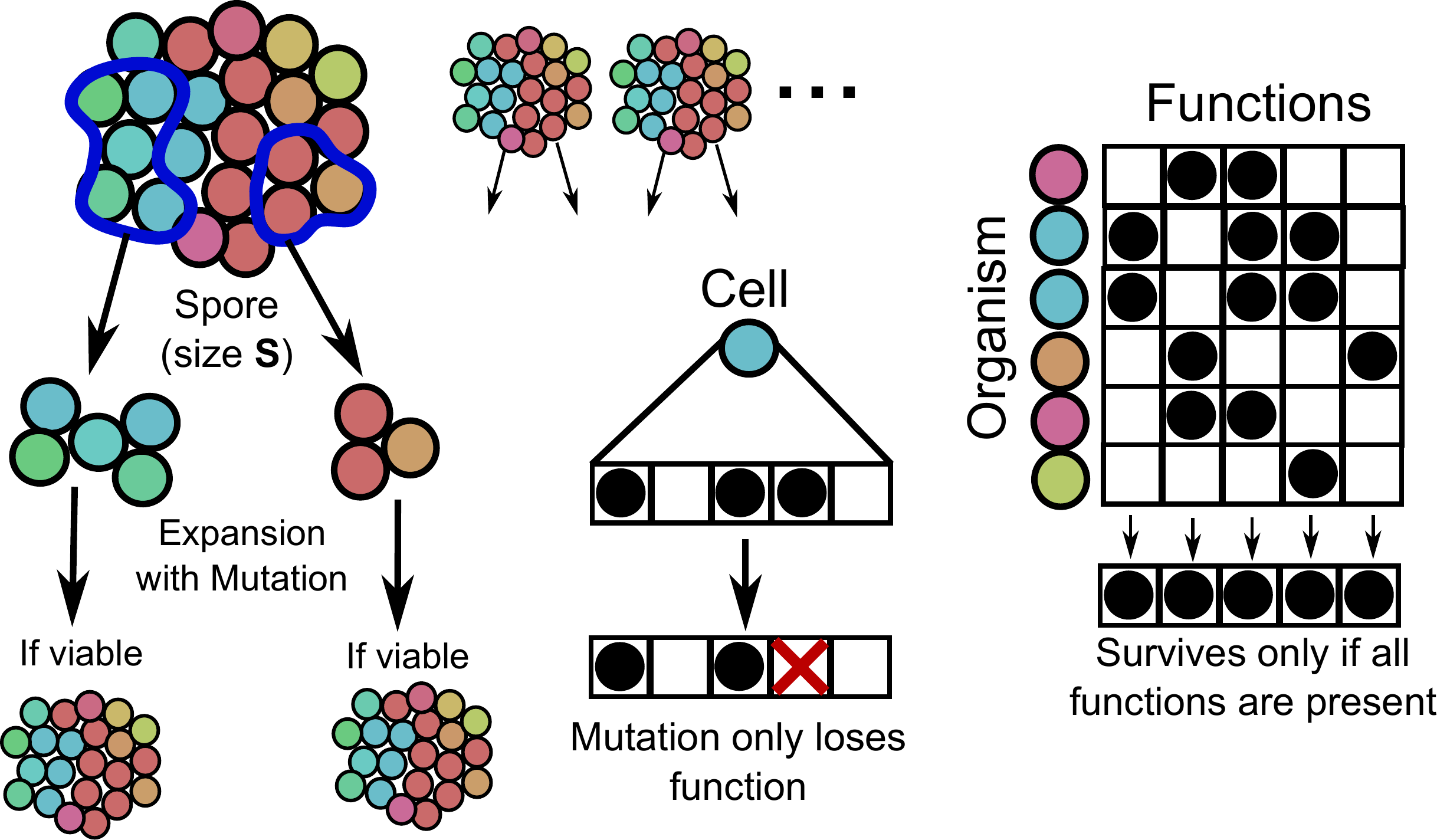}
\caption{Schematic representation of the multi-level branching process underlying the population dynamics. Each cell, represented by circles of different colors, has a number of potential functions (white boxes), which are filled with black circles if the cell possesses that function. When cells replicate, they can mutate and lose functions. The multicolored clusters of cells represent biofilms. When a biofilm replicates, a subset of the cells that comprise it is selected. The size of these subsets is dependent on a variable property associated to each cell whose distribution at the level of the biofilm may fluctuate between biofilms and through generations. This subset is tested for its ability to collectively represent a full complement of the potential functions. If it does, it proceeds to seed a new biofilm and the individuals that make it up replicate with loss-of-function mutation until the carrying capacity of an individual biofilm is reached. In each generation, each biofilm is given a certain number of opportunities to replicate with different spores of different sizes. If the biofilm-level carrying capacity is reached, the replication process stops. In some cases, the population may not reach the biofilm-level carrying capacity, and thus, there is potential for global extinction.
}
\label{fig:modelschema}
\end{figure*}

In this section we provide a general presentation of the model and its analysis. The lowest-level elements of the model are $N_F$ genes representing functions required for the survival of the biofilm~(Fig.~\ref{fig:modelschema}). An individual cell is comprised of a collection of these genes where, in this model, a cell may only possess at most one copy of each gene. Because of this, the genotype of each cell may be represented by a string of $N_F$ bits, where a 1 indicates the presence of the gene for function and a 0 indicates the absence of that gene. The top level of the model is a population of isolated communities (biofilms), each composed of a population of cells of fixed size $N_C$, which in turn each have a collection of $N_F$ genes encoding functions.

The biofilm-level population proceeds via cycles of replication. Every iteration, the biofilms may each generate two propagules in order to attempt to produce viable offspring, where viability is determined by having at least one copy of each of the $N_F$ genes among its cell population. Viable offspring are accumulated in a list which then becomes the biofilm-level population for the next iteration. To generate this list, parent biofilms are sampled randomly without replacement until either every biofilm has attempted two replications, or until the list has reached a maximum carrying capacity $N_P$. This mechanism allows the biofilm population size to fluctuate and even to go extinct.

The biofilms themselves reproduce by randomly selecting a subset of their cells (with replacement) to generate a propagule of a sporesize $N_S$ (this does not remove the parent cells from the community). The propagule is then expanded through replication of its cells. The replication process is modelled by initializing the film with the propagule, then sampling randomly from all cells in the film with replacement to choose the parent of the next cell. This is repeated until the biofilm has grown to size $N_C$. Then, the biofilm is evaluated for viability: if the film contains all necessary functions, it is viable and replaces another random biofilm in the population. Otherwise, it is discarded.

The replication process of the individual cells involves two evolutionary operators: deleterious mutation and horizontal gene transfer. During a replication event, each $1$ in the bit string may be lost to mutation with rate $m$ (changing the bit from a $1$ to a $0$). In addition, with rate $h$ the value of that bit may be replaced with the corresponding value possessed by another cell randomly sampled from the same biofilm at that point in time. Horizontal gene transfer in this model transmits both 0 and 1 values equally, based on their presence in the population.

For much of the analysis of this model, we consider the sporesize $N_S$ to be fixed and directly investigate the survival rate and population structure at that sporesize. However, in order to investigate the evolutionary dynamics we must also specify where the information about the sporesize is stored within the biofilm. If we choose to have it be stored at the biofilm level, then there is a clear way for the biofilm-level individuals to assert top-down control over their components. As such, it seems better to associated the sporesize with individual cells instead. To do this, we also associate an integer, SS, to each cell which is its preferred sporesize, and which mutates via randomly increasing or decreasing with probability $m_S$ per replication. The first cell that is chosen to create a new propagule determines the size of that propagule based on its preference. This means that a parasite could improve its chance of survival by asking for a very large propagule. This increases the chance of including non-parasitic cells. Because we are investigating the conditions that allow for the evolution of the multicellular state, we choose this mechanism to make the stability of the multicellular state as unlikely as possible. In this way, if the multicellular state still emerges, then this can be considered to be a more robust feature of the model. In actual biological systems, this constraint may be applied, if at all, in a less stringent manner.

\section*{Results}

\subsection*{Fixed Sporesize}

We refer to the version of our model where sporesize is allowed to fluctuate as dynamic. First we study the case in which the sporesize is fixed. We do this to determine the structure of the evolutionary landscape on which the dynamic version of the model will operate, to help guide our understanding of what is happening microscopically. The aim of this is to study the underlying structure induced by mutation and the replicative bottleneck, without regard to extra factors such as different replication rates among cells within the film due, for example, to the metabolic load deriving from the requirement of producing the macromolecules necessary to perform particular functions. These factors may be added in later, but would obscure the structure that arises purely from genetic drift combined with the replicative bottleneck.

As a proxy for fitness, we measure the fraction of a biofilm's offspring that are viable. Survivability in our model is generally determined by the ability of a community of cells to possess a collection of genes. In some cases, the community contains only one cell type, in which case that cell would be required to possess a copy of each gene. If the community contains multiple cell types, it is possible for the community to be composed of cells that collectively possess at least one copy of each gene. This quantity is not stationary, but in fact changes as mutations accumulate in the population over time. If we start with a biofilm in which every cell has all functions, the survival rate is initially very high but begins to decay as functions are lost, until it reaches a point where selection pressure at the biofilm level prevents it from decaying further. As such, the landscape actually experienced by a biofilm depends on the speed of its evolution. 

At short times, larger sporesize always corresponds to higher survivability. This is because as more cells are transferred, there is a much larger chance of capturing all necessary functions in the child film. The survival curve in this regime is very flat when the population is initially homogeneous and function-complete, as non-viable films require every cell in the spore to spontaneously develop the same deleterious mutation. As such, the death rate is exponentially suppressed as the sporesize grows, and a very large spore is not actually needed to ensure survival.

As time passes, the population of cells within the biofilms begin to accumulate mutations. On the one hand, if the sporesize is equal to the biofilm size, mutations do not risk causing the death of the offspring until the biofilm approaches the point where each function is only present in approximately one of its cells. At that point, however, it is possible for a biofilm to give rise to a sterile offspring --- equal to the probability that any of the needed functions is lost to a deleterious mutation when the cells replicate. On the other hand, if the sporesize is 1, then any cell with any deleterious mutation will give rise to a non-viable offspring if it is picked to be the spore. As such, in that limit the death rate is just the probability of a random cell in the film having at least one deleterious mutation. However, because of that, deleterious mutations cannot be accumulated --- all cells that give rise to viable offspring are identical. The degree to which the single-cell spore has a lower death rate than the biofilm-sized spore is determined by the number of functions $N_F$ which must be retained.

\begin{figure}
\includegraphics[width=\columnwidth]{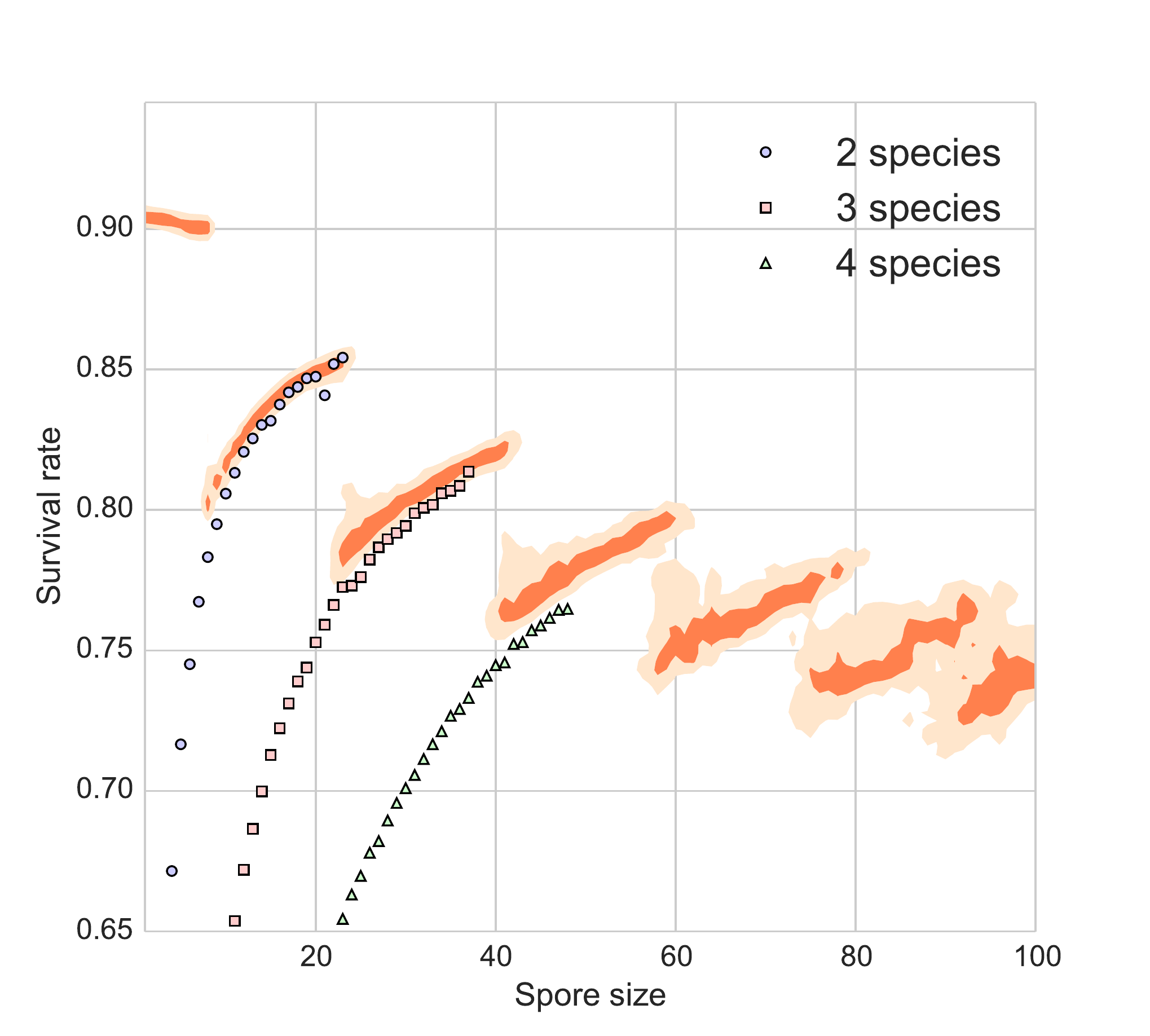}
\caption{Histogram of survival rate after 5000 generations for evolutions
at constant spore size. The color contours show the
distribution of outcomes for 86 simulations. The circles,
squares and triangles correspond to cases where 2, 3 and
4 species have been used to initialize the population.}
\label{fig1}
\end{figure}

Between the two extreme cases, simulations of the model ($N_F=10$, $N_C=100$, $N_P=100$, $m=0.01$, $h=0$, 5000 generations) show that there is a sequence of collapses where the survivability decreases in steps (Fig.~\ref{fig1}), but locally increases with increasing sporesize. This can be understood by thinking about the minimal number of cells needed to construct a viable spore. If only one cell is needed, then as the sporesize increases the probability of finding that cell also increases (and so the survival rate increases). However, if the sporesize becomes too large, non-viable cells begin to be included. Once this effect grows to the point that the minimal viable offspring requires two different cells from the parent (e.g. no single cell in the film has all functions), the large spore becomes only half as effective at guaranteeing their inclusion. This proceeds in sequence as an increasing number of distinct cells are required to make a viable offspring. The discreteness of the apparent landscape is due to the discreteness of the number of required cells, and so as the number of required cells becomes larger the steps in the landscape become more blurred.

\begin{figure}
\includegraphics[width=\columnwidth]{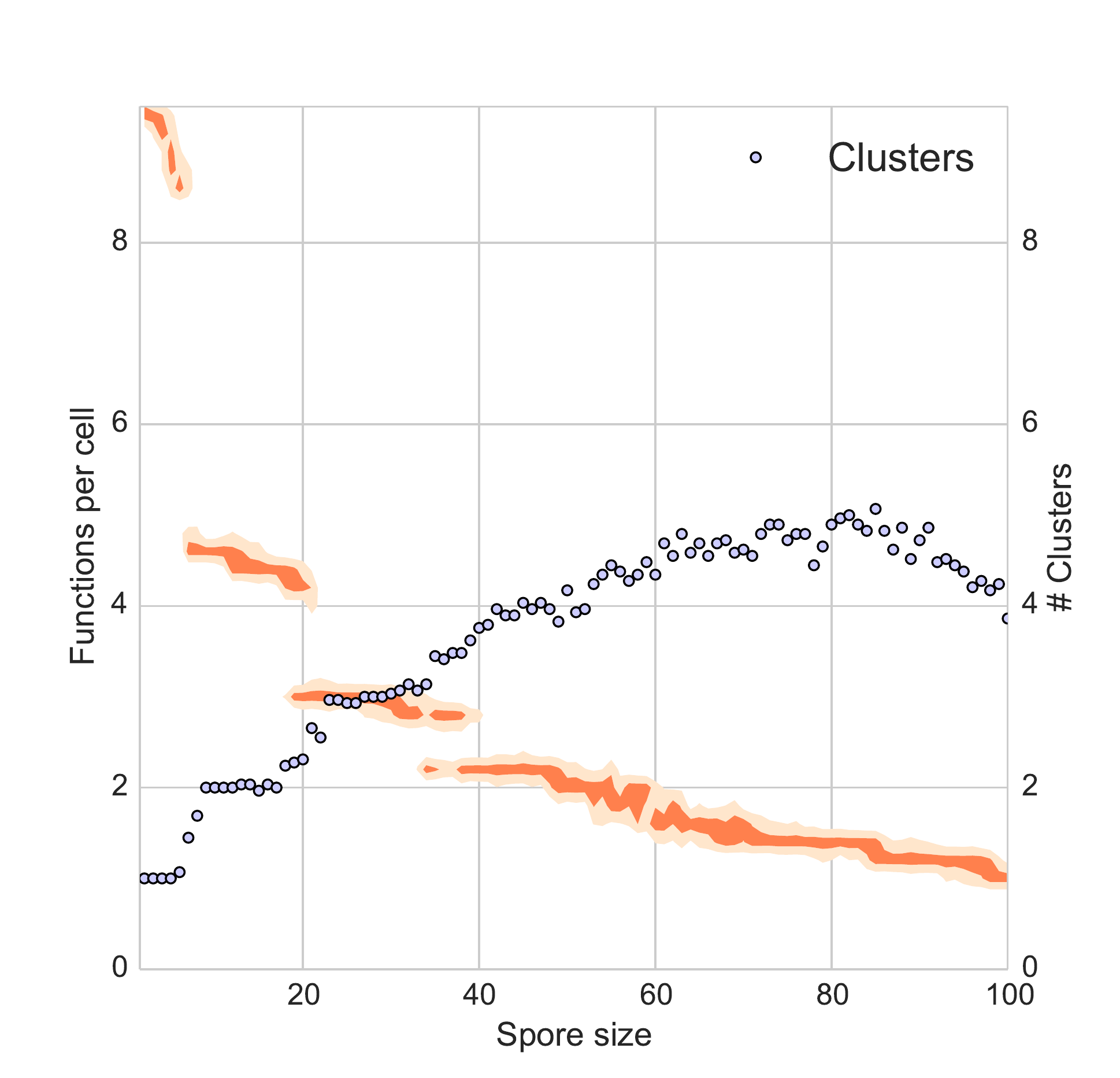}
\caption{Histogram of the average number of functions per cell, plotted against the number of clusters in sequence space detected using DBSCAN ($\epsilon=0.05$, min\_samples=1000),
averaged over 29 simulations. There are discrete jumps in the number of clusters with increasing sporesize as the population structure changes, corresponding to a division of necessary functions across multiple member species.}
\label{fig2}
\end{figure}

We test this idea by preparing populations that are pre-configured to have two, three, and four required sub-populations for making a viable offspring. For example, the two-species case has half of the initial cells containing only the first $5$ functions with the other half having only the last $5$ functions, and so on. These are the marked lines plotted against the histogram data in Fig.~\ref{fig1}. We note that the marked data points continuously join up with the different discrete segments of the histogram data, which supports the idea that the sporesize is controlling the strength of population-level repair. Furthermore, we use the DBSCAN algorithm \cite{ester1996density,pedregosa2011scikit} to cluster the population of individual cells in sequence space. This gives us an estimate of the number of distinct species of cell as a function of sporesize (Fig.~\ref{fig2}). We see well-defined discrete steps in the number of discovered clusters for the lower-half of the range of possible spore sizes. These correspond to jumps in the death rate.

\subsection*{Fluctuating Sporesize}

If the only genetic mechanism in the system is deleterious mutation, there is no possibility to recover lost functions or to have internal repair of the population. In that case, the only repair mechanism available is selection, which for large spore sizes only becomes significant once the population has already suffered a build-up of harmful mutations. After this build-up leading to a damaged population the biofilm can no longer survive with a smaller sporesize. There is thus a ratchet effect which may lock a biofilm population into a particular large sporesize at long times. The local gradient of the fitness is always indicating a preference for larger sporesize, so the system cannot easily detect the sudden jumps in survivability that would be obtained from smaller sporesize in order to find the fitness optimum. In addition, taking advantage of those jumps from the state of a less-structured population is impossible if the only genetic operators are deleterious mutation, as once there are no cells in the biofilm that have multiple functions a multi-function cell can no longer be recreated. This raises the question of accessibility of the survival rate landscape for an actual dynamical system.

\begin{figure}
\includegraphics[width=\columnwidth]{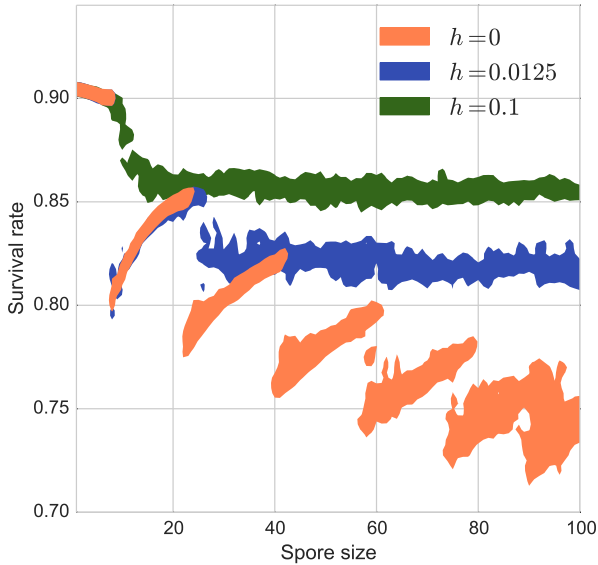}
\caption{Survivability for fixed sporesize with different levels of horizontal gene transfer ($m=0.01$, $N_F=10$, $N_C=100$, $N_P=100$,  5000 generations). The presence of even a small amount of horizontal gene transfer allows the population structure of the biofilm to rebuild itself, smoothing out the cliffs in the survivability landscape.}
\label{fig3}
\end{figure}

In the presence of a horizontal gene transfer mechanism the pool of available functions within a biofilm may be redistributed between its members. This means that mutational loss of population structure is not permanent in a given biofilm lineage, which causes the sharp transitions between population structures to become blurred (Fig.~\ref{fig3}). As a result, a small amount of horizontal gene transfer may prevent the system from getting stuck and enable the system to transition to smaller sporesizes and towards multi-function cells. However, even though the optimum survivability is at a small sporesize, it remains to be seen whether or not the system will discover this optimum under its own evolutionary dynamics when starting far from it. 

\begin{figure}
\includegraphics[width=\columnwidth]{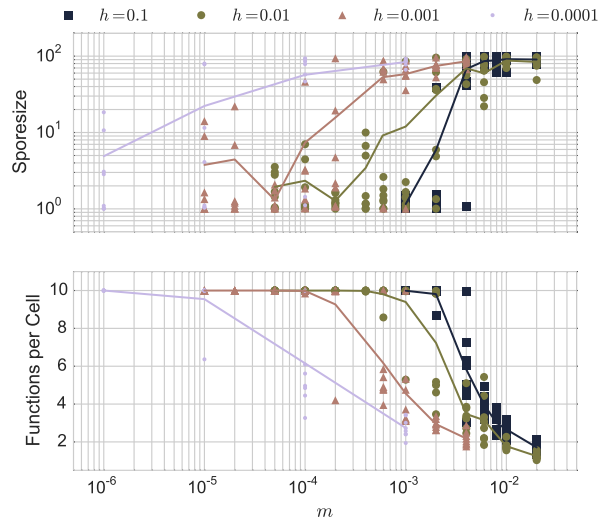}
\caption{Behavior of the model when the sporesize is allowed to change for different mutation rates and HGT rates ($m_s = 0.01$, $N_F=10$,$N_C=100$, $N_P=100$, $2.5\times10^6$ generations). At long times, there is a first order phase transition between the initial phase with sporesize comparable to the biofilm size and $~1$ function per cell, and the multicellular phase with small sporesize and $~N_F$ functions per cell. }
\label{fig4}
\end{figure}

To test this, we initialize the system in a state where each cell only has one function, and allow the preferred sporesize of each cell to vary via a random walk whenever the cell replicates, with probability $m_s$. We observe that in the case of non-zero horizontal gene transfer, there is a first order phase transition with respect to mutation rate $m$ \cite{Goldenfeld1992}. The transition occurs from the initial state, which is characterized by a sporesize comparable to the biofilm size and approximately one function per cell, to a state with $O(1)$ sporesize and cells containing close to a full complement of functions (Fig.~\ref{fig4}). The signature of the first order nature of the phase transition is an extended coexistence regime in which there is bistability between the two phases. The larger the horizontal gene transfer rate, the larger the mutation rate at which the phase transition occurs.

\begin{figure}
\includegraphics[width=\columnwidth]{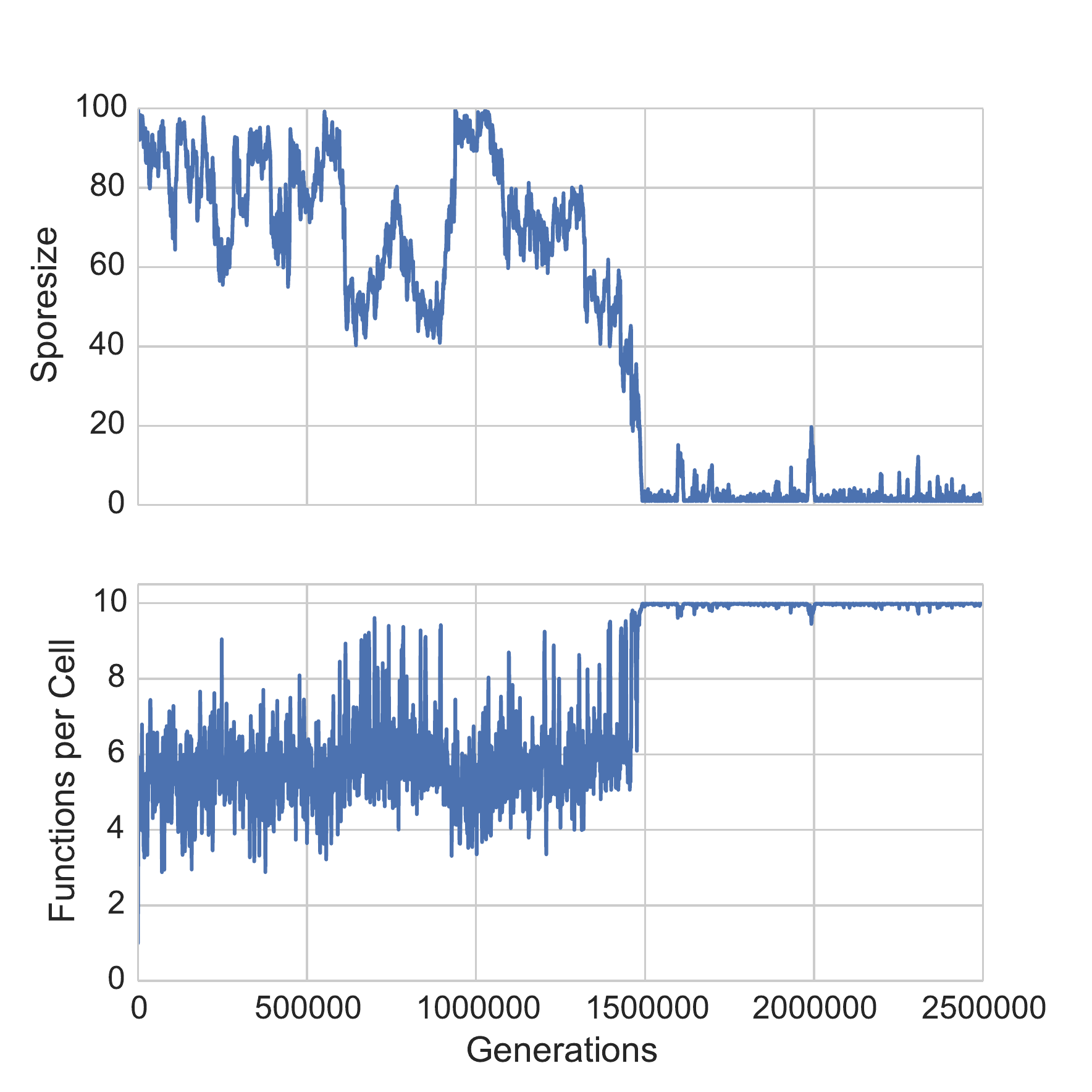}
\caption{Example evolutionary trajectory for $m=0.001$, $h=0.01$, $m_s=0.01$, $N_F=10$, $N_C=100$, $N_P=100$, showing the dynamics of the average sporesize and average number of functions per cell.}
\label{fig5}
\end{figure}

An example evolutionary trajectory from the bistable regime ($m=0.001$, $h=0.01$, $N_F=10$, $N_C=100$, $N_P=100$) is plotted in Fig.~\ref{fig5}. The system does not progressively evolve towards the multicellular state, but rather spends most of its time fluctuating in distinct phases. The development of the multicellular state is rapid when it finally begins, but must be nucleated by a relatively rare event.

\section*{Discussion}

An important feature of our study is that the global increase in survivability is associated with long-time, discontinuous changes in population structure. This makes the evolution of multicellularity particularly non-trivial. We have shown that horizontal gene transfer provides one mechanism for resolving this dilemma, as it is capable of smoothing the fitness landscape over sporesize thereby carving a favorable path from uni- to multicellularity. This kind of mechanism enables the population structure to better navigate the space of possible states.

The mechanism of a replicative bottleneck gives rise to evolutionary forces which act to repair variations in population structure and to transfer selective pressure from the biofilm scale to the cell scale. This allows a system to create robust organization on a larger scale, in the form of a genetically homogeneous colony replicating through a single cell gamete: a multicellular individual. In a more general sense, the mathematics of this process can be mapped onto other systems in which evolutionary dynamics act on multiple scales --- for example, the genetic complement of a protocell and the unification of multiple shorter genes into a single chromosome.

Our study suggests that the natural methodology to think of this transition is that of the Eigen error-threshold \cite{eigen1989molecular}. There is some mutation rate which drives the population to forget the optimal structure (e.g., each cell having each function). As the population structure decays, the selection pressure resisting each further deleterious mutation begins to increase until it reaches some stable point which sits just at the error threshold. The smaller the sporesize, the more quickly the selection pressure increases with response to population decay, and so the closer to the `complete' population the system ends up being. Because requiring $n$ cells versus requiring $n+1$ cells is a very large difference in survivability when $n$ is small, the system tends to cluster around the edges of those states as that is where the error threshold changes most rapidly. This observation may provide valuable insights into experiments.

The model we have investigated here is intended to act as a baseline or `minimal model' for the phenomenon of the emergence of a higher level of organization. However, there are still certain elements of the model which admit variations that we have not investigated here. The reason for excluding these extra factors is that we want to focus on the costs and benefits specifically associated with the creation of a narrow replicative bottleneck. Different versions of the dynamics, along with alterations to the survivability of the biofilm or the replication rates of the component cells can be easily added on top of this basic model in order to extend it or to calibrate it to specific experimental systems. For the same reason, we do not consider the effects of metabolic costs associated with encapsulating several functions within individual cells, which could lead to different replication rates among individuals. 

\section*{Summary}

In the presence of multiscale selection and a variable information bottleneck, selection at the collective scale interacts with the bottleneck limiting information flow from the individual cell scale. When this bottleneck is very wide, the effect of selection is weak, but as the bottleneck shrinks this causes an amplification of selection pressure and in turn stabilizes the population at the small scale. The result is a feedback which causes the bottleneck to either expand to minimize the cost of selection and effectively reducing the impact of large-scale patterns, or to shrink as far as possible and so allowing selective forces operating on the collective scale to be transmitted to the cell level. This decision of whether to run off towards large bottlenecks or small bottlenecks is a phase transition in the evolutionary dynamics of the system. The new form of order corresponding to this phase transition is the emergence of a new unit of well-defined individual at the collective scale.

We have characterized via a minimal model of a three-layer system that it is possible to observe, \emph{in silico}, a transition possessing several of the most salient features of the evolutionary transition from unicellularity to multicellularity. Two evolutionary forces co-exist to shape the structure of the population and stabilize the emergence of the higher-level structures associated with multicellularity. We have argued that this is due to the existence of a replicative bottleneck. The transition has to go through a series of population changes that locally reduce survivability and we have shown that mechanisms such as horizontal gene transfer help smooth these jumps and ease the process of global evolution towards multicellularity. Further work will include more detailed formulation of potential experimental tests for these predictions and the study of chained transitions that may serve as a model analogous to the occurrence of more than one of the major transitions in evolution within the context of a single dynamical model \cite{MaynardSmith1995}.

\section*{Acknowledgments}
We would like to thank Aaron Dinner, Christoph Adami, Nigel Goldenfeld, and Nicholas Chia for useful discussions regarding the development of the model and connections to different types of biofilm behaviors, Tetsuya Yomo for useful discussions regarding the connection to evolutionary dynamics in protocells, Raymond S. Puzio for discussions regarding mathematical analysis of the model, and Norman Packard for suggestions towards improving the presentation of the manuscript.

\end{document}